\documentclass[sigconf]{acmart}

\def\BibTeX{{\rm B\kern-.05em{\sc i\kern-.025em b}\kern-.08emT\kern-.1667em\lower.7ex\hbox{E}\kern-.125emX}}
    
\renewcommand\footnotetextcopyrightpermission[1]{} % removes footnote with conference information in first column
\makeatletter
\renewcommand\@formatdoi[1]{\ignorespaces}
\makeatother
\begin{document}

%
% The "title" command has an optional parameter, allowing the author to define a "short title" to be used in page headers.
\title[Recommender system for prompting omitted foods in dietary assessment surveys]{Validation of a recommender system for prompting omitted foods in online dietary assessment surveys}

%
% The "author" command and its associated commands are used to define the authors and their affiliations.
% Of note is the shared affiliation of the first two authors, and the "authornote" and "authornotemark" commands
% used to denote shared contribution to the research.
\author{Timur Osadchiy}
\email{t.osadchiy@newcastle.ac.uk}
\orcid{0000-0003-2174-6250}
\affiliation{%
  \institution{Open Lab, School of Computing, Newcastle University}
  \streetaddress{Urban Sciences Building, 1 Science Square, Science Central}
  \city{Newcastle upon Tyne}
  \country{United Kingdom}
  \postcode{NE4 5TG}
}

\author{Ivan Poliakov}
\email{ivan.poliakov@newcastle.ac.uk}
\affiliation{%
  \institution{Open Lab, School of Computing, Newcastle University}
  \streetaddress{Urban Sciences Building, 1 Science Square, Science Central}
  \city{Newcastle upon Tyne}
  \country{United Kingdom}
  \postcode{NE4 5TG}
}

\author{Patrick Olivier}
\email{patrick.olivier@monash.edu}
\affiliation{%
  \institution{Community Org \& Social Informatics, Monash University}
  \city{Melbourne}
  \country{Australia}
}
 
\author{Maisie Rowland}
\email{maisie.rowland@newcastle.ac.uk}
\affiliation{%
  \institution{Institute of Health and Society, Newcastle University}
  \streetaddress{Baddiley-Clark Building, Richardson Road}
  \city{Newcastle upon Tyne}
  \country{United Kingdom}
  \postcode{NE2 4AX}
}

\author{Emma Foster}
\email{emma.foster@newcastle.ac.uk}
\affiliation{%
  \institution{Institute of Health and Society, Newcastle University}
  \streetaddress{Baddiley-Clark Building, Richardson Road}
  \city{Newcastle upon Tyne}
  \country{United Kingdom}
  \postcode{NE2 4AX}
}

%
% By default, the full list of authors will be used in the page headers. Often, this list is too long, and will overlap
% other information printed in the page headers. This command allows the author to define a more concise list
% of authors' names for this purpose.
\renewcommand{\shortauthors}{Osadchiy, et al.}

%
% The abstract is a short summary of the work to be presented in the article.
\begin{abstract}
Recall assistance methods are among the key aspects that improve the accuracy of online dietary assessment surveys. These methods still mainly rely on experience of trained interviewers with nutritional background, but data driven approaches could improve cost-efficiency and scalability of automated dietary assessment. We evaluated the effectiveness of a recommender algorithm developed for an online dietary assessment system called Intake24, that automates the multiple-pass 24-hour recall method. The recommender builds a model of eating behavior from recalls collected in past surveys. Based on foods they have already selected, the model is used to remind respondents of associated foods that they may have omitted to report. The performance of prompts generated by the model was compared to that of prompts hand-coded by nutritionists in two dietary studies. The results of our studies demonstrate that the recommender system is able to capture a higher number of foods omitted by respondents of online dietary surveys than prompts hand-coded by nutritionists. However, the considerably lower precision of generated prompts indicates an opportunity for further improvement of the system.
\end{abstract}

%   <concept>
%     <concept_id>10003120.10003121.10003122</concept_id>
%     <concept_desc>Human-centered computing~HCI design and evaluation methods</concept_desc>
%     <concept_significance>500</concept_significance>
%   </concept>

% \ccsdesc[500]{Human-centered computing~HCI design and evaluation methods}

%
% Keywords. The author(s) should pick words that accurately describe the work being
% presented. Separate the keywords with commas.

%
% This command processes the author and affiliation and title information and builds
% the first part of the formatted document.
\maketitle

\section{Introduction}

Dietary assessment is a pivotal part of healthcare systems. Accurately capturing dietary habits and measuring intake is essential for analyzing the role of diet in causing and preventing a range of chronic diseases including diabetes, cardiovascular disease and cancers. The reliability of assessment methods directly affects the accuracy of targeting of public health policies and interventions. The use of conventional in-depth interviewer-led surveys that heavily rely on labor of nutritionists and dietitians fails to scale and keep up with a rapidly increasing population. This has resulted in an ongoing endeavor of various research teams to computerize paper-based dietary assessment tools (e.g. food frequency questionnaires and 24-hour recalls) \cite{carter2015development,simpson2017iterative,subar2012automated}. Such instruments equip dietary assessment with flexibility, wider coverage and remote collection of responses.

Unfortunately, automated dietary assessment methods, as their interviewer-led paper-based predecessors, are heavily dependent on memory of a respondent and prone to omissions. Respondents may skip reporting associated foods (e.g. milk in coffee/tea, butter on toast) due to forgetting them, insufficient training or their impression that reporting those details is not important. Some respondents may also choose to alter their intake or leave out certain foods (e.g. fast food, alcohol) to avoid social desirability bias \cite{shim2014dietary}. Social desirability bias is the tendency of an individual to report an answer in a way that will be viewed favorably by others (i.e. socially acceptable) \cite{hebert1995social}. Such omissions can be identified and minimized by a trained interviewer who can inquire for more details when needed. Thus, to generate reliable and accurate results an intelligent dietary assessment system has to emulate this behavior of an interviewer and adapt to the diet of an individual. That system could apply data driven approaches, such as approaches used in recommendation systems, to previous responses of respondents to learn about their eating habits and provide appropriate cues for the memory of interviewees.

Recommender systems are widely used in various domains for inferring user preferences, from entertainment to online retail \cite{covington2016deep,linden2003amazon}. For example, recommender systems in online shops analyze users' purchasing histories and behavior to suggest items relevant for specific needs of an individual \cite{linden2003amazon}. Similarly, a dietary assessment system can use data driven methods to learn about the eating behavior of respondents and prompt them in relation to foods they have potentially eaten but omitted when recalling their dietary intake. 

In our previous work we developed a recommender system based on pairwise association rules that aids respondent's recall \cite{osadchiy2019recommender}. The main contribution of this paper is an evaluation of the recommender system in online dietary assessment surveys, in real-life settings. Further, we provide the background of systems for conducting online dietary assessment surveys. We review the development of a recommender algorithm for one of such systems. We then report and discuss the design and results of two dietary surveys, in order to compare the efficiency of the recommender to that of prompts added by nutritionists. 

\section{Previous work}
\subsection{Challenges in dietary assessment}
Dietary assessment methods can be broadly split into three categories: nutrient biomarkers, objective (i.e. direct observations) and subjective (i.e. self-reported intake by subjects) \cite{shim2014dietary,zuniga2015considerations}. Biomarkers can be used to estimate intake of some nutrients (e.g. fatty acids) and energy \cite{arab2003biomarkers,bingham2002biomarkers}. Previous research has demonstrated a number of advantages of biomarkers, including high accuracy, the lack of social desirability bias and an independence from a subject's ability to self-report their intake \cite{potischman2003biologic}. Nevertheless, for most nutritional biomarkers, this method imposes a number of practical and economical challenges including the need to collect, store and analyze blood, urine or other biological specimens \cite{thompson2010need}. Biomarkers are still highly useful in calibrating measurement errors in dietary reports. For example, doubly labeled water is widely considered as a gold standard that other dietary assessment methods are compared with \cite{livingstone2003markers,bingham2002biomarkers,burrows2010systematic}.  Another approach to collecting data for dietary assessment involves skilled research staff directly observing and recording a subject's intake \cite{shim2014dietary}. Records are collected in a subject's home environment and include not only information about food consumption but also preparation methods. This method is especially useful in developing countries, with subjects with a low literacy rate or where food is prepared in large quantities for a group (e.g. family). However, direct observations are somewhat impractical on larger scale populations.

Subjective dietary assessment methods, based on intake self-reported by subjects, includes weighted food diary method, food frequency questionnaire (FFQ) and 24-hour recall \cite{shim2014dietary}. The weighted food diary method requires subjects to weigh portion sizes and record their intake after every eating occasion. To collect accurate records this method requires subjects to undertake training before taking part in a study. Thus, the weighted food diary requires a high level of motivation and poses a relatively large burden on subjects \cite{macdiarmid1998assessing}. The method is known to change diets of subjects taking part in a study, which limits the accuracy of results \cite{rebro1998effect}. Meanwhile, the FFQ enables capturing of an individual's intake over a long period of time (e.g. a month of a year) in a relatively simple, cost- and time- efficient manner \cite{shim2014dietary,thompson2010need}. FFQ is an advanced form of a checklist that normally contains 100-150 foods and asks a subject to report which of those foods they have eaten over a specific period, and how much \cite{shim2014dietary}. Answers for the questionnaire can be both collected through an interview or via a self-administered approach (e.g. online survey). In contrast to the weighted food diary the FFQ poses a lower subject burden \cite{thompson2010need}. The accuracy of this method, however, is relatively low as it relies on the ability of subjects to remember their diet over a long period of time and thus, is prone to misreporting \cite{kristal2005time,thompson2010need}. 

One of the most widely adopted approaches is the multiple-pass 24-hour recall, which is considered to offer a more favorable balance of high accuracy and low subject burden \cite{johnson2002dietary}. 24-hour recall was designed as an interviewer-led method to collect information about all foods and drinks consumed by an individual for a previous day \cite{guenther1996multiple}. Capturing habitual intake using this method requires multiple non-consequent interviews to be conducted over a long period of time (e.g. weeks, months). The estimation of energy intake with this method is relatively accurate when validated with doubly labeled water \cite{lopes2016misreport}. However, similarly to the FFQ, it is also prone to under-reporting due its reliance on subjects self-reporting their intake \cite{thompson2010need}. 

\subsection{Online dietary assessment surveys}
The involvement of skilled professionals in the collection and analysis of complex dietary data has both economic implications and implications for scalability. To address these issues a number of systems have been developed that replace an interviewer in the 24-hour recall method with an online survey \cite{carter2015development,simpson2017iterative,subar2012automated}. For researchers, such systems provide tools to administer surveys and collect dietary records from participants with a detailed breakdown of the foods and drinks consumed, to enable the estimation of energy intake and intakes of macro- and micro-nutrients. Respondents are asked to go through a survey in a form of a web-based interface and record their dietary intake for the previous day. The survey comprises a series of questions about each meal and all its constituent foods and drinks. The collected information includes names of foods/drinks and their portion sizes. Portion sizes are self-estimated by a respondent using validated photographs of weighted servings of foods \cite{foster2010development}. Each respondent normally records their meals for the previous day on three separate occasions. A single day (i.e. recall submission) typically consists of four to seven meals (e.g. breakfast, morning snack, lunch, evening meal etc.). Each reported meal may include a selection of foods, drinks, desserts, condiments and such (referred to generically as foods).

Participants commonly omit foods and under-estimate portion sizes in dietary assessment surveys due to a range of human factors including poor human memory or lack of attention \cite{macdiarmid1998assessing,thompson2010need}. A study of 83 adults between 20 and 60 years old demonstrated that interviewer-led 24-hour recall underestimated energy intake on average by 33\% of that measured with doubly labeled water \cite{lopes2016misreport}. Various state-of-the-art techniques have been proposed by the research community to aid respondents' self-reported intake recall and minimize under-estimation. For example, images of food captured by respondents using handheld devices or wearable cameras can enhance the quality of self-reported dietary surveys by revealing unreported foods and misreporting errors \cite{gemming2015image,ahmad2016mobile}. In addition to that, image recognition algorithms applied to those images may facilitate the accuracy of energy and portion size estimation \cite{zhu2015multiple,liu2016deepfood}. Another direction for improving the accuracy of dietary assessment is monitoring eating bahaviour using wearable devices and sensors embedded into the environment \cite{kalantarian2017survey}. However, probably the most widely adopted method of reminding respondents about omitted foods in online dietary assessment surveys are associated food questions \cite{subar2012automated,bradley2016comparison}. Associated food questions are questions about foods that are commonly consumed together. Normally each associated food question is manually recorded into the system's database by a nutritionist or a dietitian along with a corresponding relation between an antecedent food (e.g. white bread, toast) and a consequent food (e.g. butter) \cite{osadchiy2019recommender}. The system returns an associated food prompt once a respondent reports an antecedent food. However, identifying relations for thousands of foods stored in the system seems impractical. Moreover, even an experienced nutritionist may not always be able to identify all the factors and aspects of a diet of a specific population.

\subsection{Recommender system}
Eating habits depend on a range of factors, including region, culture and a specific diet, which makes them hard to predict. Thus, manual extraction of associated food questions and keeping them up-to-date is prone to omissions and is a time-consuming task. Those challenges motivated us to develop a recommender system that extracts food associations in an automated manner \cite{osadchiy2019recommender}. The recommender algorithm was developed for Intake24, open-source dietary assessment system for conducting large-scale population 24-hour multiple-pass recall surveys online \cite{simpson2017iterative}.

While developing the recommender system we considered a few existing approaches such as collaborative filtering. Collaborative filtering is, for example, used in online retail websites including Amazon, to recommend products that customers might want to buy based on their purchase history \cite{linden2003amazon}. This approach requires a system of ratings, so customers can express their attitude to products. The system then uses collaborative filtering to build a model of user preferences based on ratings that users assign to products. That model can then produce recommendations based on similarity between users or products in terms of their preferences and correlations in ratings. However, in online dietary assessment surveys respondents will likely only ever use the system if they are a part of a study and only for a short period of time. Therefore, it is unlikely that the system will aggregate personal behavioral profiles rich enough for applying collaborative filtering or a similar method, which is referred as the cold-start-problem \cite{lam2008addressing}. For that reason, for the implementation of the associated foods recommender system we looked into other methods that are independent of personal user profiles. Instead, we selected a method which built a model of eating behavior of a population \cite{osadchiy2019recommender}. The model is then used to generate prompts based on foods selected by an individual during their recall. In our previous paper we considered three approaches to the development of the recommender system based on: (i) association rules \cite{dumouchel2001empirical}, (ii) implicit social graph \cite{roth2010suggesting} and (iii) pairwise association rules \cite{osadchiy2019recommender}. We compared the three versions of the recommender system in an evaluation on a large data set of real dietary recalls \cite{osadchiy2019recommender}. The evaluation has demonstrated that the implementation based on pairwise association rules performs better for the defined task \cite{osadchiy2019recommender}.

To build a model of eating behavior the recommender system based on pairwise association rules takes a dataset of all meals reported by a given population, where a meal is a group of foods (e.g. egg, toast, butter, cheese, orange juice) reported in a single intake. Each meal is split into unique pairs of foods (e.g. toast and butter, toast and cheese, toast and orange juice, etc.). For every pair the resulting model stores the number of meals that contain this pair. The model also stores the number of meals that contain each individual food. To produce a recommendation the algorithm takes foods reported by the current user to find pairs that contain those foods. Foods from the filtered pairs that have not been reported by the user yet are used as recommendations. The list of recommended foods is sorted by their likelihood of being observed in a meal having observed the reported foods in descending order. In other words, the algorithm recommends foods that are more likely to be observed with any of reported foods in pairs. 

The likelihood for a recommended food $f$ is calculated as $R_f= C_f \times W_f$, where $C_f$ is the aggregation of conditional probabilities of observing that food in a pair with one of the reported foods $f_i$ in a meal having observed that reported food; and $W_f$ is the weight of the aggregation. $C_f$ is calculated as $C_f = \sum_{i=0}^{N} \frac{count(f, f_i)}{count(f_i)} $, where $N$ is the number of reported foods, $count(f, f_i)$ is the number of meals in the dataset that contain the pair $(f, f_i)$ and $count(f_i)$ is the number of meals in the dataset that contain the reported food $f_i$ individually. $W_f$ is calculated as $W_f = \sum_{i=0}^{K} count(f_i)$, where $K$ is the number of reported foods $f_i$, for which a pair containing the recommended food $f$ being considered has been found. Thus, the reported foods that were previously observed with the recommended food and that were reported more often in the dataset give a higher weight to the recommendation. 

Previously we analyzed the performance and demonstrated the effectiveness of the recommender system over food associations hand-coded by nutritionists in a simulation of respondents omitting foods with data collected from past real-life dietary surveys \cite{osadchiy2019recommender}. In this paper we deploy and compare the recommender system to the hand-coded food associations in conditions of two real dietary surveys.

\section{Methods}
\subsection{Interface design}
\begin{figure}[h] 
  \centering    
  \includegraphics[width=0.7\linewidth]{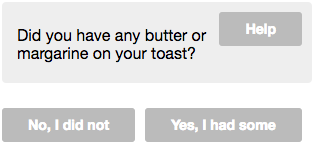}
  \caption{Hand-coded associated food prompt in Intake24.}
  \Description{Hand-coded associated food prompt in Intake24.}
  \label{fig:hand-coded-prompt}
\end{figure}

\begin{figure}[h] 
  \centering    
  \includegraphics[width=0.7\linewidth]{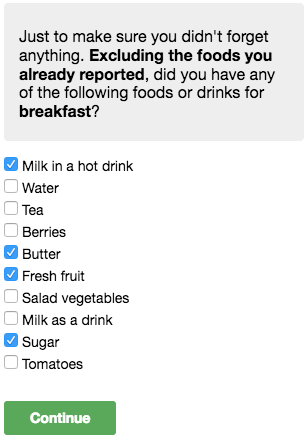}
  \caption{Generated associated food prompt in Intake24.}
  \Description{Generated associated food prompt in Intake24.}
  \label{fig:generated-prompt}
\end{figure}

The existing associated food prompts that are based on links between two foods manually added into Intake24 are returned immediately after one of the foods has been reported (Fig.~\ref{fig:hand-coded-prompt}). So, a respondent is typically prompted with multiple foods while reporting a single meal. In contrast to that, recommendations produced by the recommender system are based on a selection of foods and have a form of a list. For that reason, we designed a screen with generated food prompts in the form of a checkbox list that is returned at the end of reporting each meal (Fig.~\ref{fig:generated-prompt}). A respondent can accept multiple foods as with the hand-coded prompts. A list of recommendations is limited to 15 foods, which is a slightly larger number than the number of search results displayed to most users by online search websites \cite{burges2005learning,van2009using}.

\subsection{Recruitment and procedure}
\subsubsection{First study.}
To recruit participants for the first study we disseminated an advertisement describing the purpose of the study and a web form to register interest via the University internal email system. Individuals had to complete the web form to volunteer for the study and were asked to circulate the invitation to their relatives and friends for participation. The first page of the registration form presented more details about the study and an informed consent form, where all conditions had to be accepted to be able to register. As part of the registration process we asked potential participants to provide their contact details (e.g. name, email, phone number and an age range). This study aimed to follow the original procedure that was used in previous validations of Intake24 \cite{bradley2016comparison,rowland2016field} and asked participants to record their intake in the morning. To help respondents to follow the procedure we queried their preferred morning times to receive reminders about completing a dietary recall. To be able to take part in this study individuals had to be over 18 years old, speak English, have a diet that is common for the UK. To minimize the likelihood of variations in diets of respondents affecting the results of the study we informed them that it is important to avoid changing their diet during the study. We assume that completing a survey may have a different complexity across devices with various screen sizes and input methods (e.g. touch interface, keyboard). For that reason we encouraged respondents to use the same device to record their intake during the study. Participants were informed that they can withdraw from the study at any time. For completing five dietary recalls in five days participants were offered a \pounds30 Amazon voucher. Recruitment resulted in 50 participants, of whom $n=26$ were men and $n=24$ were women. One participant chose to withdraw during the study. The participant's ages ranged between 18 and 64 years.

The study was conducted over a three-week period with a separate group of participants in each week. We asked participants to record their intake for five consecutive days from Monday to Friday. Every morning participants received automated email and SMS reminders to log onto Intake24 and report the meals they had for the previous day. Participants were asked not to record their meals elsewhere (e.g. notepads) to aid their recalls. During the first three days (Monday - Wednesday) the system presented one type of associated food prompts and in the remaining two days (Thursday - Friday) the other type of prompts. We presented hand-coded food prompts in the first three days to $n=19$ participants. For $n=30$ participants we presented hand-coded food prompts in the last two days. Monday recall submissions were used as an introduction to minimize the learning effect by familiarizing participants with the system's interface and the 24-recall procedure. Those recalls were discarded during the analysis of the results and only diet recalls of work days (Monday - Thursday) were used.

\subsubsection{Second study.}
The second study was collected during the Newcastle Can campaign that aims to help to reduce obesity levels in the North East of UK \cite{newcastleCan2018}. The campaign used various methods to motivate its participants to improve their diets. One of the methods involved participants recording their meals for a previous day using Intake24 to receive feedback on their diet and links to NHS web pages with more information about healthy eating.

To access Intake24, participants had to first click a corresponding button in their Newcastle Can personal profile page. Before they were redirected to Intake24 a modal window was displayed informing participants that they are transferred to an external website and by proceeding they accept the privacy policy of Intake24. The privacy policy informs a reader that anonymized and aggregated information may be subject to processing for scientific purposes and used as a basis for publications. All the recalls from Newcastle Can users were anonymous and did not contain any personally identifiable information.

Compared to the first study, the second did not pose any specific requirements to the time of a day or days of the week, when participants needed to record their intake. The study did not require respondents to use any specific device or to submit a certain number of recalls either. In this study we collected recalls with hand-coded prompts for a week and recalls with generated prompts for another week. The second study attracted $n=91$ respondents to complete at least one recall, of whom $n=77$ were women and $n=14$ were men. The age of participants in this study ranged from 18 to 82 years.

\subsection{Statistical analysis}
We measure the proportion of recalls with generated and hand-coded food prompts, where respondents accepted at least one food. We also analyze the mean number of foods accepted per recall in these two settings. In addition, we measure precision of food prompts, that is the number of accepted foods divided by the number of returned foods. Many participants who reject associated food prompts that are still relevant for other participants genuinely believe that they reported all foods. For example, although someone may drink coffee without milk and sugar, a food prompt querying about milk and sugar in a coffee is still relevant for many. For that reason, we analyze the mean number of accepted foods only for recalls, where at least one prompted food was accepted. We compare the coverage of two types of prompts through the number of unique foods that were returned and accepted. We also compare the estimates of energy reported with two types of prompts. We exclude recall submissions with abnormally low-calorie content from the analysis (below 250 kcal for the whole day). To detect potential changes in usability of the survey interface we compare the mean duration of recalls, i.e. time it took respondents to complete a survey. We assume that longer recalls may indicate that food prompts generated by the recommender system negatively affected the usability of the interface. Previous research shows that participants complete their recalls using Intake24 in 16 minutes on average \cite{rowland2016field}. For that reason, while analyzing the mean duration of recalls we ignore those that took longer than 60 minutes, since that could indicate that respondents took a break while completing their recall. To analyze the significance between two given means we use Mann-Whitney U test.

\subsection{Results}

The first study resulted in 96 submitted recalls with hand-coded and 97 with model-generated food prompts. In the survey with hand-coded prompts 69\% of respondents used a desktop computer, 30\% used a mobile device, and 1\% used a tablet. In the survey with generated prompts, 69\% of respondents used a desktop computer, 28\% used a mobile device, and 3\% used a tablet. Hand-coded food prompts were displayed at least once in 86 recalls and accepted in 57 recalls (66\%). Generated prompts were returned in 97 recalls at least once and accepted in 61 recalls (63\%). For the hand-coded prompts that is 1.1 accepted food per recalls on average, whereas for the generated prompts that is 2.3 foods, which is significantly higher (P < 0.001). Precision of the hand-coded and generated food prompts for the first study are 24\% and 2\% respectively.

\begin{figure}[h] 
  \centering    
  \includegraphics[width=\linewidth]{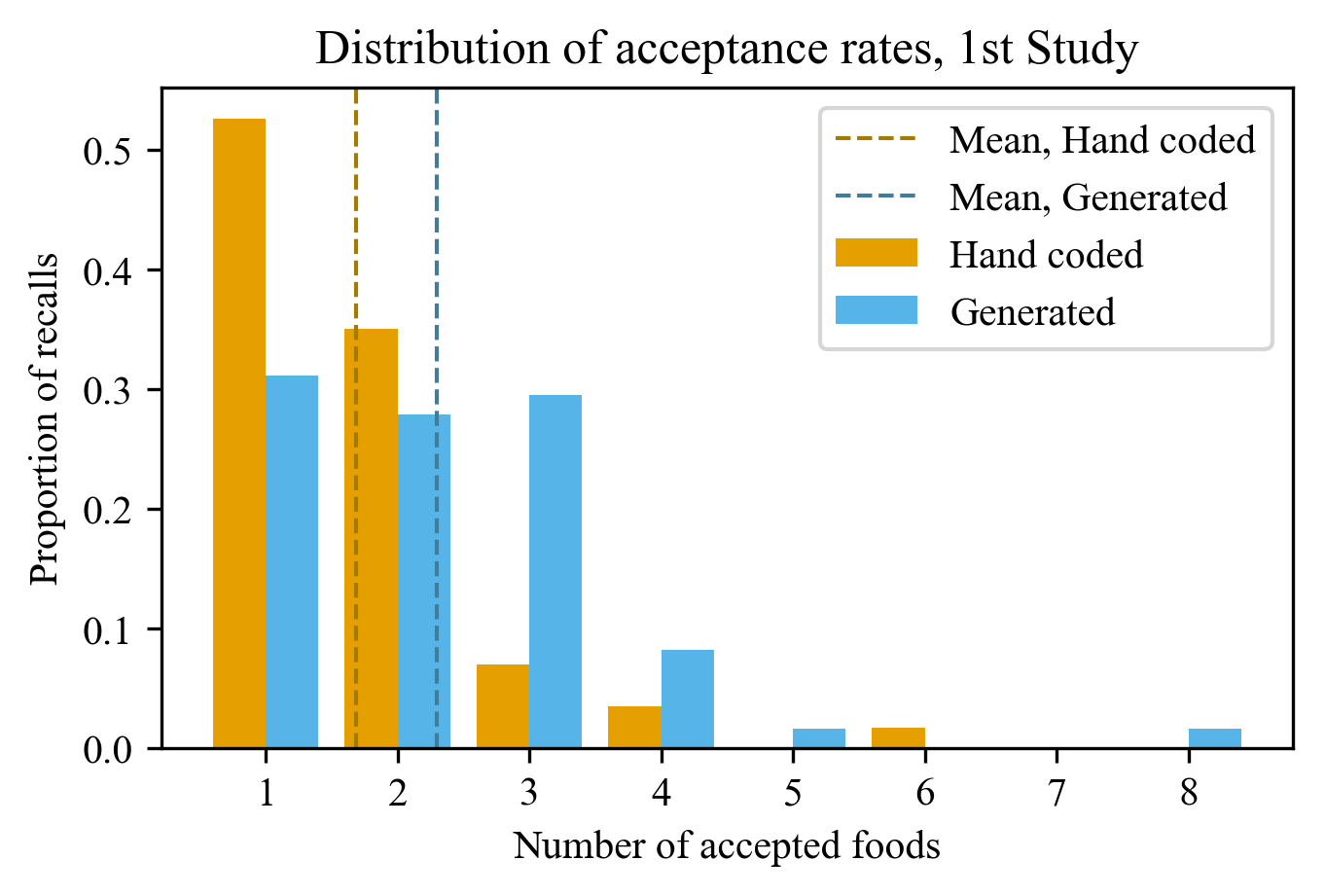}
  \caption{Distribution of accepted foods per recall during the first survey.}
  \Description{Distribution of accepted foods per recall during the first survey.}
  \label{fig:acceptance_1}
\end{figure}

\begin{figure}[h] 
  \centering    
  \includegraphics[width=\linewidth]{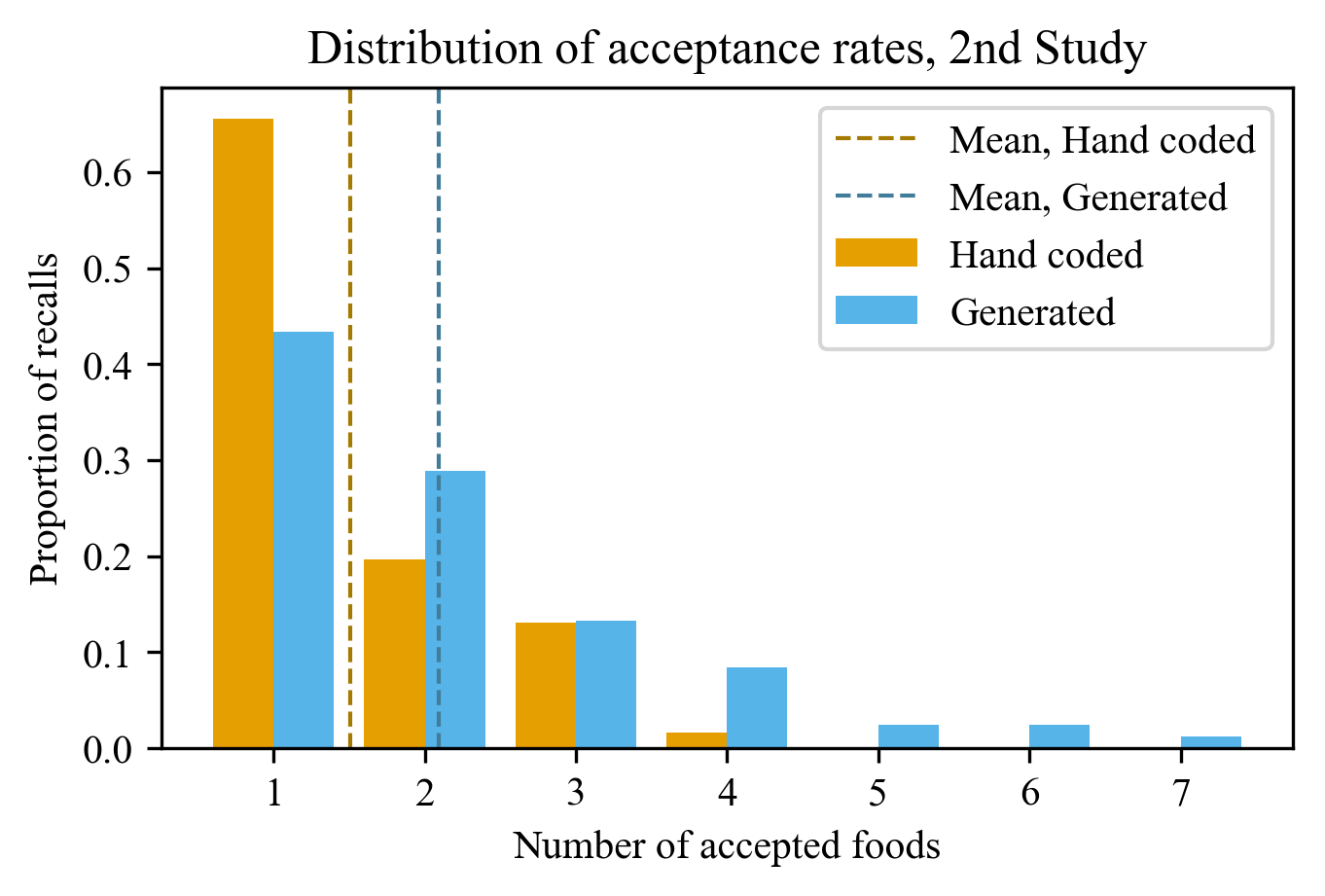}
  \caption{Distribution of accepted foods per recall during the second survey.}
  \Description{Distribution of accepted foods per recall during the second survey.}
  \label{fig:acceptance_2}
\end{figure}

Participants in the second study submitted 133 recalls with hand-coded and 119 with model-generated food prompts, of which 41 (31\%) and 57 (48\%) recalls respectively were submitted between Saturday morning and Monday night. Intake records assisted by the hand-coded prompts were submitted on average between 1:20pm and 9pm. Similarly, recalls assisted by the generated prompts were submitted on between 1:25pm and 8:30pm. In surveys with hand-coded prompts, 37\% of respondents used a mobile device, 33\% used a desktop, and 30\% used a tablet. In surveys with generated prompts, 50\% of respondents used a mobile device, 22\% used a desktop, and 28\% used a tablet. Hand-coded food prompts were displayed at least once in 122 recalls and accepted in 61 recalls (50\%). Generated prompts were returned in 115 recalls and accepted in 83 recalls (72\%). As in the first study, there is a significant difference between the mean acceptance rates. Participants on average accepted 1.5 foods per recall from hand-coded prompts and 2.1 from generated prompts (P = 0.002). Precision of the hand-coded and generated food prompts for the second study are 16\% and 2\% respectively. Histograms of acceptance rates for both types of prompts in two studies are provided in Fig.~\ref{fig:acceptance_1},~\ref{fig:acceptance_2}.

\begin{figure}[h] 
  \centering    
  \includegraphics[width=\linewidth]{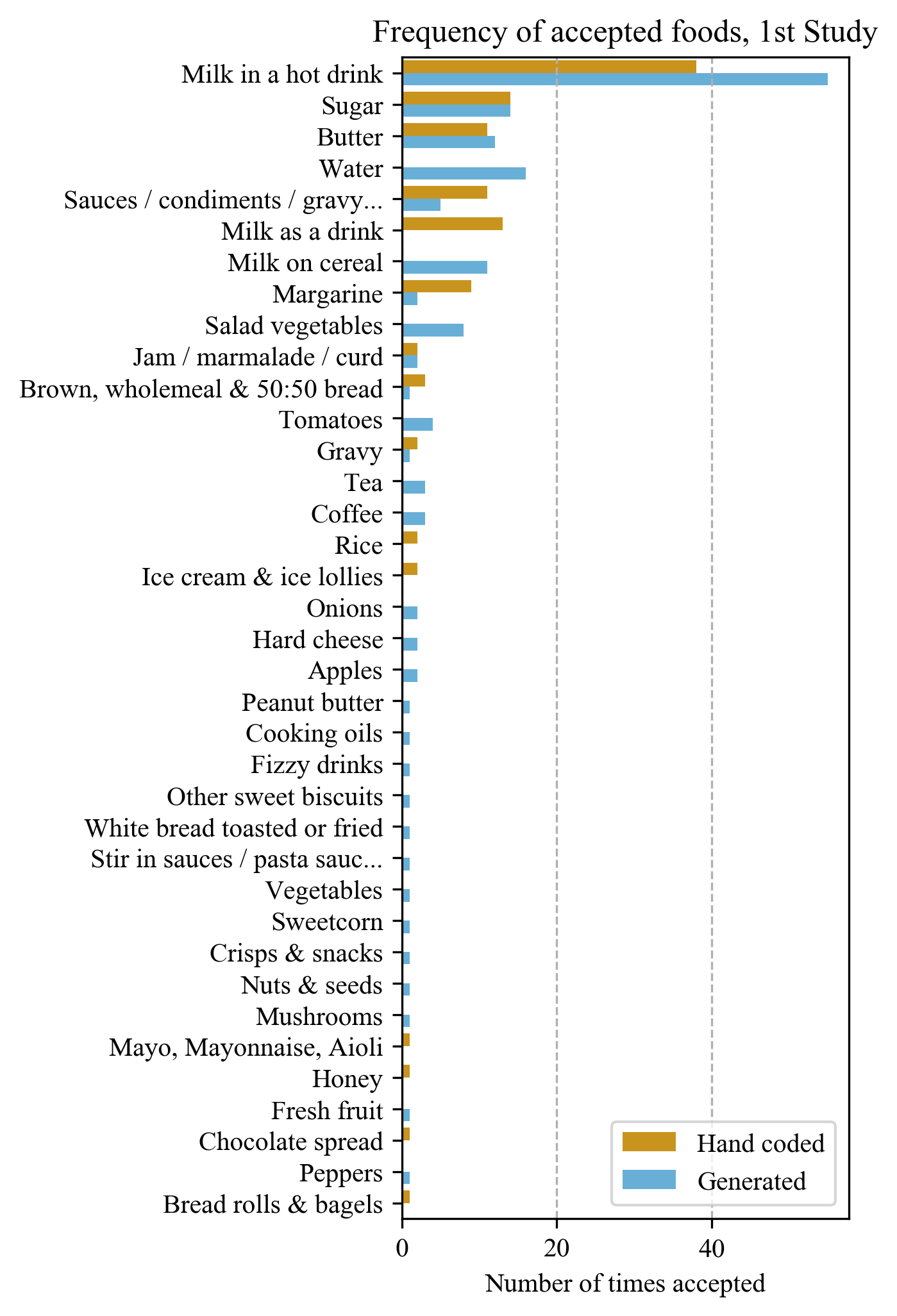}
  \caption{Frequency of accepted foods during the first second survey.}
  \Description{Frequency of accepted foods during the first second survey.}
  \label{fig:food_1}
\end{figure}

\begin{figure}[h] 
  \centering    
  \includegraphics[width=\linewidth]{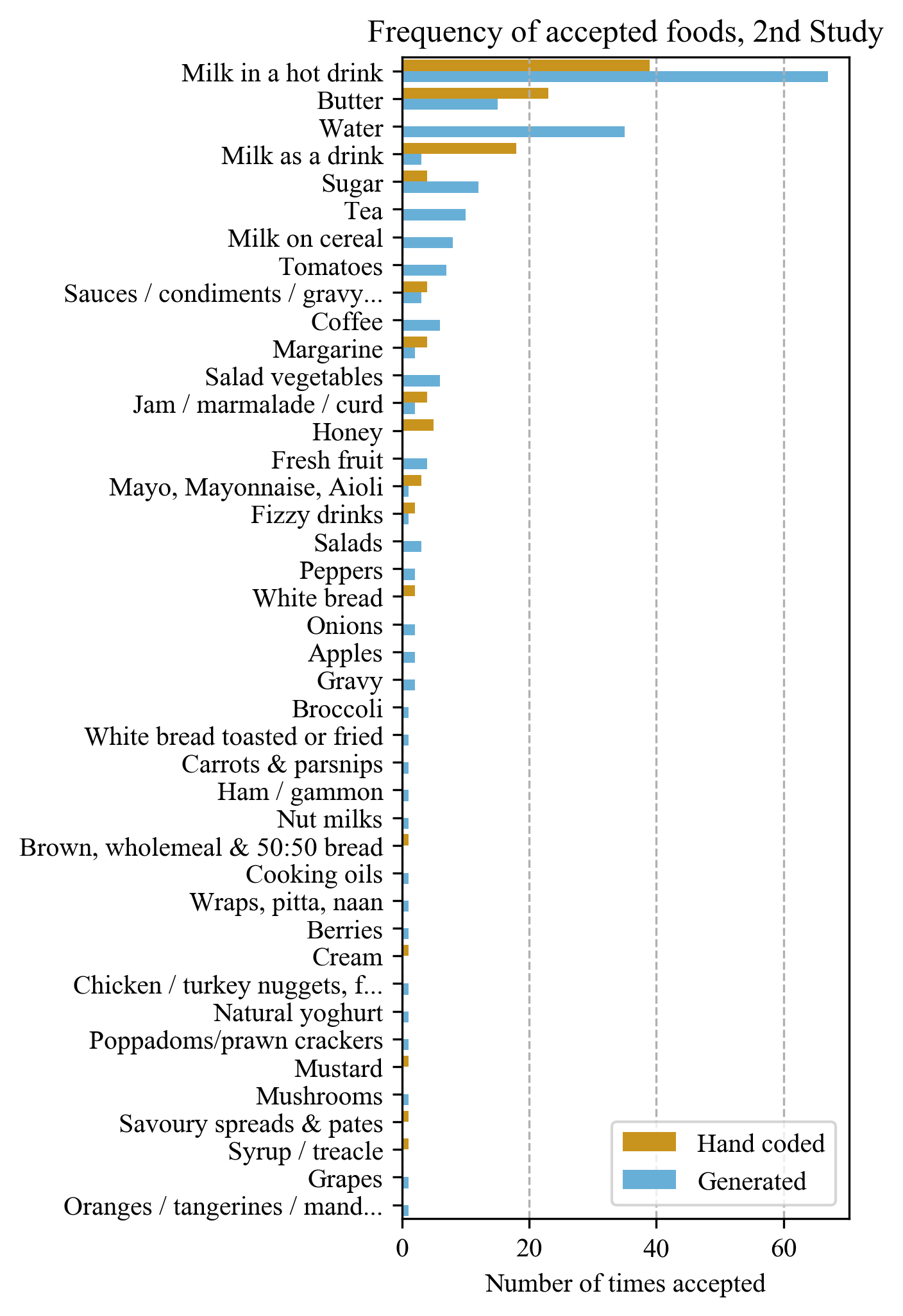}
  \caption{Frequency of accepted foods during the second survey.}
  \Description{Frequency of accepted foods during the second survey.}
  \label{fig:food_2}
\end{figure}

In the first and the second studies with hand-coded prompts the number of unique foods accepted by participants was 15 (9\%) and 16 (9\%) out of 164 and 186 unique reported foods respectively (Fig.~\ref{fig:food_1},~\ref{fig:food_2}). In surveys with generated prompts, this number was found to be at least twice as high with 30 (18\%) and 35 (19\%) out of 165 and 189 unique reported foods (Fig.~\ref{fig:food_1},~\ref{fig:food_2}).

We found no significant difference between the mean energy reported in recalls with hand-coded food prompts (1911.8 kcal) and generated food prompts (1790.6 kcal) during the first survey (P = 0.159). However, the effect was observed during the second survey, where the mean reported energy for hand-coded prompts was 1461.7 kcal and 1545.7 kcal for generated prompts (P = 0.02).

While examining the mean duration of recalls, 7 (4\%) recalls were excluded from the first survey and 15 (6\%) from the second survey because these were longer than 60 minutes. We observed no significant differences in the duration of recalls in either survey. The mean duration of recalls during the first survey was 15.9 minutes for recalls with hand-coded prompts and 13.3 minutes for recalls with generated prompts (P = 0.108). During the second survey the mean duration of recalls was 15.9 and 16.3 minutes respectively (P = 0.297).

\section{Discussion}
\subsection{Principal findings}
The results show that associated food prompts generated by the recommender system based on pairwise association rules have been demonstrated to be an effective and scalable alternative to prompts hand-coded by nutritionists in online dietary assessment surveys. This supports findings from our previous research \cite{osadchiy2019recommender}. In the meantime, according to the recommendations from the NHS UK a man needs around 2,500 kcal a day to maintain his weight given a healthy and a balanced diet \cite{calorie2019nhs}. For a woman, that figure is around 2,000 kcal a day. Compared to that, the mean energy intake estimated in the first study during the current research is considerably low, which suggests that there were cases of omissions and under-estimation. This is in agreement with previous research that found Intake24 to underestimate energy intake by 1\% on average compared to the interviewer-led recall \cite{bradley2016comparison}; and that the interviewer-led recall  underestimates energy intake by 33\% on average compared to that measured with doubly-labeled water \cite{lopes2016misreport}. The second study was conducted as part of a weight loss campaign, which could explain even lower energy estimates. At the same time, participants of the second study reported higher energy intake in recalls assisted by generated food prompts, which may indicate an improved accuracy of assessment.

Meanwhile, generated food prompts have shown significantly lower precision compared to that of hand-coded prompts. This is due to a much longer list of recommendations produced by the system. As it is observed in this study, only 30 and 35 unique foods that had been recommended by the system were accepted across all recalls in the first and second surveys. This is in agreement with previous studies that some foods are more likely to be omitted \cite{bradley2016comparison}. This also explains a considerably higher performance of the recommender in a simulation we conducted in the previous study, where we assumed that any food can be omitted \cite{osadchiy2019recommender}. To shorten the list of recommendations, we plan to incorporate the acceptance rate of recommended foods into the recommender system. This will potentially allow us to place foods that are more likely to be forgotten higher up the list of recommendations.

The mean time it took respondents to complete a survey remained similar for generated and hand-coded prompts. The observed proportion of desktop and touch devices used in recalls assisted with hand-coded and generated prompts is comparable in both studies. In the first study, respondents submitted their recalls in the morning. In the second study, time of the day when respondents recorded their intake is similar in both conditions. Thus, relatively stable duration of recalls assisted with hand-coded and generated prompts may indicate that the usability of the system was not affected by the longer list of generated food prompts.

We should note that two types of prompts have a different presentation format. Hand-coded prompts are presented as questions that can be accepted or rejected by a respondent during reporting a meal. Generated prompts are presented in a form of a list after the meal has been reported. Thus, the observed greater number of accepted food items from the the recommender system may be an effect of the presentation format rather than an effect of the better fit of the suggested food items. In addition to that, the recommender system was trained on data collected in the past surveys, where hand-coded associated food prompts were used. Therefore, there is a chance that would there be no hand-coded prompts, respondents might have reported some foods associations less often or not reported them at all. Hence, the recommender system potentially would not pick those associations from the data in the training process. Nevertheless, in our two studies, more than a half of foods captured by the recommender system were not defined in the database of hand-coded prompts. Thus, the two methods of prompting could potentially complete each other at least when the system has been deployed for a new population and there is no representative data for training the recommender system yet.

\subsection{Limitations}
The current studies involved a relatively small number of participants, and the recruitment method used in the first study meant that the demographics profile of our respondents was limited which may limit the applicability of the results. Furthermore, in the second study, participants were members of a weight loss campaign which may imply that they are more health-conscious than the general population. Finally, the age and gender distribution of participants was not balanced in both studies. Thus, further research needed to validate our findings using a wider range and balanced distribution of demographical backgrounds.

In the first survey participants did not submit recalls for Friday and weekend days, when people commonly consume more energy and diet is less structured than on the week days. That is somewhat addressed in the second survey that included weekends. However, in the second survey participants could submit their meals at any time of a day, which contradicts with the original multiple pass 24-hour recall method, where participants are asked to record their meals in the morning. To address these concerns a similar study should follow the procedure of the 24-hour recall method offering respondents to complete their recalls for three non-consecutive work days and on one day over the weekend.

Lastly, in both studies, in cases when participants rejected food prompts, it was not possible to verify whether they genuinely reported all foods, ignored prompts or prompts did not contain any relevant foods. Similarly, when respondents accepted prompts, we have no evidence that they actually consumed those foods. In the future, meals reported by participants could be verified against direct observations.

\section{Conclusions}
In this paper we aimed to improve methods used in online dietary surveying to assist respondent's memory and improve the accuracy of dietary assessment. We designed a recommender system that prompts about foods potentially omitted by respondents during a dietary survey as an alternative to prompts hand-coded by nutritionists and dietitians \cite{osadchiy2019recommender}. In contrast to other contexts, where recommender systems are used (e.g. online retail, entertainment), dietary assessment systems are limited in their ability to collect enough data for each individual user to build a model of their personal preferences. For that reason, the recommender system described in this paper mines a model of eating behavior of a population from data collected in previous dietary surveys. To validate associated food prompts generated by the recommender system we compared their performance to prompts that were hand-coded by nutritionists in two online dietary assessment surveys. The validation demonstrated the ability of the recommender system to capture significantly larger number of foods omitted by a respondent per recall than that by the hand-coded prompts. Moreover, the number of distinct foods accepted by respondents from the generated prompts was at least twice higher than that from the hand-coded prompts in both surveys. That indicates that more than a half of omitted foods captured by the recommender system were not foreseen by nutritionists. In the meantime, to ensure that a dataset that is used for training the recommender system contains associations of foods commonly omitted by respondents it should be collected in surveys assisted by prompts hand-coded by professionals with nutritional background. Judging by the average time it took respondents of both surveys to complete their recalls we observed no difference in complexity of completing a survey with the two prompting methods. At the same time, considerably low precision of generated prompts indicates an opportunity to further improve the relevance of recommendations produced by the system. The results produced in this paper indicate the effectiveness of using the recommender system over hand-coded rules in online dietary assessment surveys, where there is a rich dataset of past surveys for training the system. 

%
% The next two lines define the bibliography style to be used, and the bibliography file.
\bibliographystyle{ACM-Reference-Format}
\bibliography{recommender-main}

\end{document}